\documentclass[Afour,sageh,times]{sagej}

\usepackage{moreverb,url}

\usepackage[colorlinks,bookmarksopen,bookmarksnumbered,citecolor=red,urlcolor=red]{hyperref}

\usepackage[table,xcdraw]{xcolor}
\usepackage{graphicx}
\usepackage[normalem]{ulem}
\usepackage{natbib}
\usepackage[T1]{fontenc}
\usepackage[utf8]{inputenc}

\definecolor{mygray}{gray}{0.9}

\usepackage{soul}
\usepackage{comment}
\usepackage{booktabs}

\newcommand\edit[1]{\textcolor{black}{ #1}} % for new additions
\newcommand\redit[1]{\textcolor{black}{ #1}} % for minor revision additions

\newcommand\BibTeX{{\rmfamily B\kern-.05em \textsc{i\kern-.025em b}\kern-.08em
T\kern-.1667em\lower.7ex\hbox{E}\kern-.125emX}}

\def\volumeyear{2016}

\begin{document}
\runninghead{Kawakami et. al.}

\title{\textit{AI Failure Loops} in Devalued Work: The Confluence of Overconfidence in AI \& Underconfidence in Worker Expertise}

\author{
Anna Kawakami\affilnum{1},
Jordan Taylor\affilnum{1},
Sarah Fox\affilnum{1},
Haiyi Zhu\affilnum{1},
Kenneth Holstein\affilnum{1}\\
\noindent\textit{Forthcoming at Big Data \& Society.}
}

\affiliation{
\affilnum{1}Human-Computer Interaction Institute, Carnegie Mellon University, USA
}

\corrauth{Anna Kawakami, Human-Computer Interaction Institute, Carnegie Mellon University, USA}

\email{akawakam@andrew.cmu.edu}

% affilaition code, anonymous for now 
% \affiliation{\affilnum{1}Sunrise Setting Ltd, UK\\
% \affilnum{2}SAGE Publications Ltd, UK}

% \corrauth{Alistair Smith, Sunrise Setting Ltd
% Brixham Laboratory,
% Freshwater Quarry,
% Brixham, Devon,
% TQ5~8BA, UK.}

% \email{alistair.smith@sunrise-setting.co.uk}

\begin{abstract}
A growing body of literature has focused on understanding and addressing workplace AI design failures. However, past work has largely overlooked the role of the devaluation of worker expertise in shaping the dynamics of AI development and deployment. In this paper, we examine the case of feminized labor: a class of devalued occupations historically misnomered as ``women's work,'' such as social work, K-12 teaching, and home healthcare. Drawing on literature on AI deployments in feminized labor contexts, we conceptualize \textit{AI Failure Loops}: a set of interwoven, socio-technical failure modes that help explain how the systemic devaluation of workers' expertise negatively impacts, and is impacted by, AI design, evaluation, and governance practices. These failures demonstrate how misjudgments on the automatability of workers' skills can lead to AI deployments that fail to bring value to workers and, instead, further diminish the visibility of workers' expertise. We discuss research and design implications for workplace AI, especially for devalued occupations.
\end{abstract}


\maketitle

\section{Introduction} 
%\anna{include 1-2 sentences on} 
\edit{Workplace AI systems are often touted by developers and deployers as tools to augment workers' abilities and improve their work experience \redit{\citep{brynjolfsson2025generative}}. In practice, however, these systems have frequently fallen short of meaningfully supporting workers} \redit{\citep{acemoglu2023can}}. In social services, AI-based decision support tools have been introduced for high-stakes decisions, only to be withdrawn following backlash from frontline workers~\citep{samant2021family}. In healthcare, clinicians ignore AI tools designed to assist in clinical decision-making~\citep{elwyn2013many,lebovitz2022engage}. In education, \edit{AI tutoring systems have added to teachers workloads and compromised their pedagogical autonomy---despite promises to reduce labor and enhance teaching}%deployments of AI-based tutoring systems have failed due to inadequate consideration of classroom teachers' needs and constraints
~\citep{holstein2017intelligent,utterberg2021intelligent,holmes2023unintended}. \edit{Across these sectors, AI systems fail to complement the situated, social, and tacit knowledge that underpins workers' expertise~\citep{lebovitz2021ai}.}  %\anna{move this upwards -> explicitly say developers, companies, researchers}

Prior research in human-computer interaction (HCI) and science and technology studies (STS) offers several explanations for these failures. Scholars have pointed to issues of poor contextual fit, where AI systems conflict with existing organizational structures and work practices~\citep{suchman1987plans,yang2019unremarkable}. Others have highlighted flawed assumptions about the nature of work, misaligned values, or inadequate governance, emphasizing how these design shortcomings intersect with broader labor and power dynamics ~\citep[e.g.,][]{forsythe1993construction,suchman1993categories,star1999ethnography, kawakamistudying2024}. \redit{However, one crucial dimension remains underexplored: \textit{occupational devaluation}, the systemic undermining of worker expertise, particularly within professions predominantly occupied by women and people of color~\citep{flores2023notions}.} \redit{In these professions, expertise is often not fully understood or explicitly recognized by others~\citep{huising2025introduction}, and devalued both in status and pay even when accounting for education and experience levels~\citep{flores2023notions}.}

How does the devaluation of worker expertise shape AI design, evaluation, and deployment? \redit{Current literature offers little insight into how this dynamic contributes to the repeated failure of workplace AI to genuinely enhance labor. \redit{This paper addresses this gap by placing occupational devaluation at the center of our analysis.} }We \redit{conduct a focused review of} academic and grey literature on AI deployments across three domains of \textit{feminized labor} \redit{in the United States}: K-12 education, home healthcare, and social work. 
 
Feminized labor has long been devalued in the U.S. %\edit{, and scholars have long discussed how different occupations grounded in feminized labor tend to share several characteristics.} 
Such work often involves \edit{emotional and} care-oriented labor~\citep{england2005emerging}, %\edit{Many occupations grounded in feminized labor have historically been}
is historically labeled as ``women's work,'' and continues to be predominantly performed by women and people of color~\citep[e.g.,][]{socialworknums,aidelow,taie2022characteristics}.
\edit{These jobs tend to be low-status within organizations, with limited worker influence over decision-making~\citep{cotter2001glass}. Workers in these fields frequently report that their experience is misunderstood by those in more powerful roles~\citep[e.g.,][]{socialwork}.}

We argue that workers \edit{engaged in devalued labor} %in societally devalued occupations 
are \textbf{especially vulnerable to flawed AI deployments}, which often rest on reductive understandings of their work. These systems not only fail to reflect the complexity of frontline labor but can also \textbf{further obscure and erode the visibility of workers' expertise}, creating a negative feedback loop that deepens occupational devaluation. Through a comparative analysis~\citep{schwandt2018case} of past scholarship on AI deployments across three feminized labor contexts (social work, home healthcare, K-12 teaching), we articulate \textit{why} and \textit{how} this dynamic emerges. 

To do so, we introduce the concept of \textit{AI Failure Loops}: recurring patterns of failure in the design, evaluation, and deployment of workplace AI systems, in which the devaluation of worker expertise leads to ineffective AI deployments—and those failed deployments, in turn, further erode recognition of workers’ skills. These loops are sustained across the AI development lifecycle, fueled by persistent \textit{over}-estimation of what AI can do and \textit{under}-recognition of the complexity and value of worker expertise. 

Some of these challenges are familiar: They echo Forsythe's observations of expert systems nearly three decades ago~\citep{forsythe1993construction}. When developing expert systems, AI developers excluded social and maintenance activities in their conceptualizations of ``work,'' leading to AI deployments that failed to reflect experts’ real-world work~\citep{forsythe1993construction}. Our paper examines how similar challenges have persisted with modern AI, with a focus on the role of \textit{occupational devaluation}. 
%occupational devaluation negatively impacts AI practices, and vice versa. 

Today, with recent waves of both AI hype and actual increases in AI capabilities, developers are targeting a rapidly expanding range of socially complex tasks associated with feminized occupations. Drawing lessons from past AI deployments in these contexts, we can expect this will \textit{further accelerate} the gross under-estimations of worker expertise and over-estimations of AI capabilities that has shaped these occupations in the past three decades. This paper responds to calls to redirect this trajectory toward \textit{pro-worker futures}~\citep[e.g.,][]{acemoglu2023can}, where worker expertise is %acknowledged and 
celebrated within AI practice and society more broadly, and workplace AI systems are designed to truly complement %and enhance 
worker capabilities. By centering occupational devaluation in our analysis, we extend these calls, highlighting the need for AI systems that not only avoid harm but actively contribute to the dignity and visibility of labor. 
%\edit{where worker expertise is acknowledged and celebrated within AI practice and society more broadly, and workplace AI systems are designed to truly complement and enhance worker capabilities.}

\section{Addressing Challenges in the Deployment of Workplace Technologies}\label{section:background_AIfail}
Researchers in STS, Computer-Supported Cooperative Work (CSCW), \edit{and Organizational Science} have long studied challenges in the deployment of workplace technologies. Workplace technologies are often designed or introduced by managers in ways that fail to account for the social contexts within which they are deployed \citep{suchman1987plans}. For instance, those in hyper-competitive workplaces \citep{orlikowski1992learning} may lack incentives to use collaborative technologies \citep{star1999ethnography}. Computing systems can also fail to be used in practice due to workers' limited visibility into systems and the system's limited visibility into workers' contexts \citep{suchman1987plans}. The design of workplace technologies is also highly political---it can prescribe certain ways of working, thus decreasing worker autonomy \citep{suchman1993categories}. In other words, workplace technologies can, in the Foucauldian sense \citep{foucault2023discipline}, discipline and punish \citep{suchman1993categories}. 

A growing body of literature has examined how so-called ``AI'' systems extend these trends and present new challenges~\citep[e.g.,][]{saxena2021framework,holmes2023unintended}. As modern AI systems expand the scope of tasks subject to automation, frontline workers find themselves performing new forms of hidden integration labor or ``patchwork''~\citep{fox2023patchwork} to bridge the gaps between what workplace AI systems purport to do and what they are actually able to do~\citep{mateescu2019ai,sendak2020real,fox2023patchwork}. \edit{Prior literature has  discussed how workers across occupations, from social workers and judges to radiologists, frequently ignore AI-based decision support tools due to inconsistencies with their decision-making practice and challenges around interpretability~\citep{lebovitz2022engage,kawakami2022improving,pruss2023ghosting}. These challenges are further exacerbated by flawed evaluations, where workplace AI systems intended to augment or automate complex human decision-making are built upon imperfect 'ground truth' labels, leading to over-claims of model performance~\citep{lebovitz2021ai}.}

AI researchers are increasingly invoking the language of ``participation'' to support more responsible technology creation~\citep{sloane2022participation}. Such calls have roots in Scandinavian efforts to democratize workplace technology development in the late 20th century \citep{muller1993participatory}. These participatory design engagements were facilitated by strong labor protections and partnerships between labor unionists and Marxist computer scientists \citep{spinuzzi2002scandinavian}. However, worker participation in AI development remains rare in practice, particularly within the United States where the socio-political conditions differ from the Nordic model of industrial democracy. Even when stakeholders are involved, workers are rarely empowered to substantially shape the design process\redit{, and familiar challenges often recur once technologies are deployed \citep{delgado2023participatory,kawakamistudying2024}}.% \anna{wrap up this paragraph: E.g., even when people are trying to engage workers, can still fail for XYZ (frame as continuation from prev paragraphs)}- conceptualizing ... rather than theorization ... power structures 

\redit{We build on existing literature on workplace technology challenges, using the \textit{systemic devaluation of worker expertise} as a lens to understand why some AI deployments fail to truly support workers, even in cases where developers intend to design worker-complementary AI. } 

%\edit{Prior critical computing research on the relationship between AI and labor has XXX. ...  We build on this prior work through our examination of the role of feminized occupational devaluation on AI deployment.}

%\anna{this sentence needs to be revised: What creates this gap between idealized forms of worker power and participation for AI and the realities faced by workers today? Our paper deepens our understanding of this, focusing on the role of occupational devaluation.}

%\anna{core gap: can maybe reiterate the core gap from the introduction, then state that we spcifically examine the gap in feminized labor contexts to help transition to the next subsection -- make sure it's clear that gap is prevalent in feminized context }

%\ken{Minor note: If posing the broad question above (``What creates this gap''), I wonder whether readers will first want to hear a bit more about what is already known about other causes behind this gap, beyond occupational devaluation. ... Update: see Jordan's Overleaf comment on these sentences. I think this captures what's needed here}
%\anna{ahh agreed, but leaving this as a todo for later for now}

\section{Devaluation of Feminized Labor} \label{background_femme}
Notions of what counts as ``work’’ and, relatedly, expertise are socially situated and have evolved across time~\citep{suchman1995making, star1999layers}. These notions are used to justify the societal and economic (de)valuation of human labor. Feminized labor---which often involves care-oriented labor like teaching, home healthcare, or childcare---provides a particularly striking example. Many forms of feminized labor were historically not viewed as ``work,'' but as acts of love~\citep{flores2023notions}. Even as feminized labor became formalized into paid occupations through social movements, workers struggled to gain legitimacy. Sociologists have theorized several explanations: Employers may justify low pay by conceptualizing feminized labor as ``acts of love'' provided out of ``goodwill'' (the ``prisoners of love’’ framework~\citep{england2005emerging}). The association of feminized labor with women may also lead to its devaluation precisely \textit{because} women themselves are devalued in society (the devaluation theory)~\citep{england2005emerging}.
%treiman1981women,england2005emerging,miller2016women}. 

Even when accounting for factors like workers' education level, occupations with a higher percentage of women pay less and have lower social status~\citep{flores2023notions}. The most female-dominated occupations in the United States still tend to involve care-based work, including home care (85\%), social work (85\%), and teaching at the primary and secondary levels (77\%)~\citep{socialworknums,aidelow,taie2022characteristics}. Workers in feminized occupations tend to have low status in their workplace, with little say in organizational decisions  impacting their day-to-day work---challenges which are exacerbated by a ``glass ceiling'' on career advancements~\citep{cotter2001glass}. These occupations often also face systemic resource shortages and high rates of worker burnout, all amid growing labor demands~\citep{england2005emerging}. 

%In the context of AI and automation, existing literature on feminized occupations has tended to trace high-level trends on the \textit{impacts} of technology deployments. %\ken{may want to adjust the wording for clarity - I'm having trouble understanding what ``trace high-level trends on the impacts of technology deployments'' means, and how it relates to the sentences that follow. Alternatively, maybe delete this summary in the first sentence and dive directly into the next sentence? i.e., ``...AI and automation, existing literature has quantitatively analyzed...''}
\edit{\redit{In the context of AI in \textit{feminized} occupations,} prior work has quantitatively analyzed labor automation trends, and found that many feminized roles---particularly those involving emotional and care-oriented labor---are projected to be among the most ``immune'' to future automation~\citep[e.g.,][]{automatedimm, filippi2023automation}. %Meanwhile, others that depend on clerical work are among the most likely to be automated. %certain types of feminized occupations---particularly those involving emotional and care-oriented labor---tend to be ``immune’’ to automation compared with others that depend on clerical work (e.g., secretaries). 
Yet, these studies  leave %describes high-level trends while leaving 
open \textit{how} day-to-day labor dynamics shape these trends, and what they mean for worker practices and AI. Seperately, a growing body of work documents AI failures in specific work contexts \redit{(see Section~\ref{section:background_AIfail})}. Our paper bridges these lines of inquiry, focusing on \redit{AI in feminized occupations in the United States.} Through case studies across multiple fields \redit{of feminized labor}, we introduce the concept of \textit{AI Failure Loops} to trace how the systemic devaluation of worker expertise both drives and is intensified by AI practices.}% \edit{We introduce the concept of \textit{AI Failure Loops} to show how the devaluation of worker expertise fuels recurring breakdowns in AI systems, \redit{across three domains of feminized labor}.}}

\section{\edit{Methods}}
\redit{To understand how the devaluation of worker expertise shapes AI practices, and vice versa, we conducted a focused review of academic and grey literature on AI deployments across three feminized occupations (social work, home healthcare, K-12 teaching), with analyses informed by our team's direct empirical experiences. In this section, we elaborate on how this experience informed our case selection (Subsection~\ref{methods_case}) and the literature we reviewed in our analysis (Subsection~\ref{methods_lit}).}

\subsection{\redit{Team Experience and Case Selection}}\label{methods_case}
\edit{We examine case studies of AI deployments across social work, home healthcare, and K-12 teaching, selected because  
our team has direct experiences collaborating with workers in each occupation. This experience---collectively including over ten years of work observing worker practices, designing interventions with workers, and speaking with leadership and managers to facilitate these collaborations---allowed us to ground our analysis of relevant literature in our own empirical material (e.g., field notes). } \edit{For example, our social work case study was informed by our prior findings from field research conducted across five U.S. public sector agencies~\citep[]{kawakami2022improving,kawakamistudying2024}. This includes experiences observing workers’ use of AI-based risk assessment tools~\citep{kawakami2022improving}, speaking with agency leadership and in-house AI developers in social work agencies~\citep[]{kawakamistudying2024,kawakami2024situate}, and designing interventions with social workers~\citep[]{kawakami2022care}. Our healthcare case study was informed by our experiences conducting interviews and design activities with home health aides, aide coordinators, and nurses to understand their perspectives around workplace AI~\citep[e.g.,][]{bartle2022second}. Our K-12 teaching case study was informed by our prior field research studying K-12 teachers' adoption and use of AI-based tutoring systems across hundreds of US-based schools and classrooms~\citep[e.g.,][]{holstein2017intelligent,Holstein2018,holstein2019}.}

\subsection{\redit{Literature Review and Analysis}}\label{methods_lit}
Our analysis is based on prior literature on AI deployments across the three case studies (social work, home healthcare, and K-12 teaching), each drawn from prior academic and grey literature (including but not limited to prior research from our team). Specifically, \edit{as we conducted a comparative case analysis~\citep{schwandt2018case} of AI deployments across these occupations, we began to notice common trends, which we explored further through a focused review of academic and grey literature.} 
\edit{We gathered and analyzed literature from computing (e.g., HCI, robotics, AIED) and domain-specific (e.g., healthcare, social work, and education) research venues.} 
\edit{We searched literature on field deployments of AI systems in workplace contexts~\citep[e.g.,][]{jorgensen2022making}, literature reviews on technologies deployed in specific workplaces~\citep[e.g.,][]{kuo2022understanding}, and syntheses of trends on the impacts of AI on workers~\citep[e.g.,][]{holmes2023unintended}. Our analysis was not limited to research on challenges from AI deployments.}
\edit{Through our initial comparative case analysis of findings across the authors, and this subsequent review of the literature, we iteratively  developed the \textit{AI Failure Loops} concept.}

\edit{We next present the concept of \textit{AI Failure Loops} (Section~\ref{sec:AIfailureloops}), then describe our three case studies (Section~\ref{sec:case_studies}) to illustrate how \textit{AI Failure Loops} unfold in practice.}

\begin{figure*}
    \centering
\includegraphics[width=\linewidth]{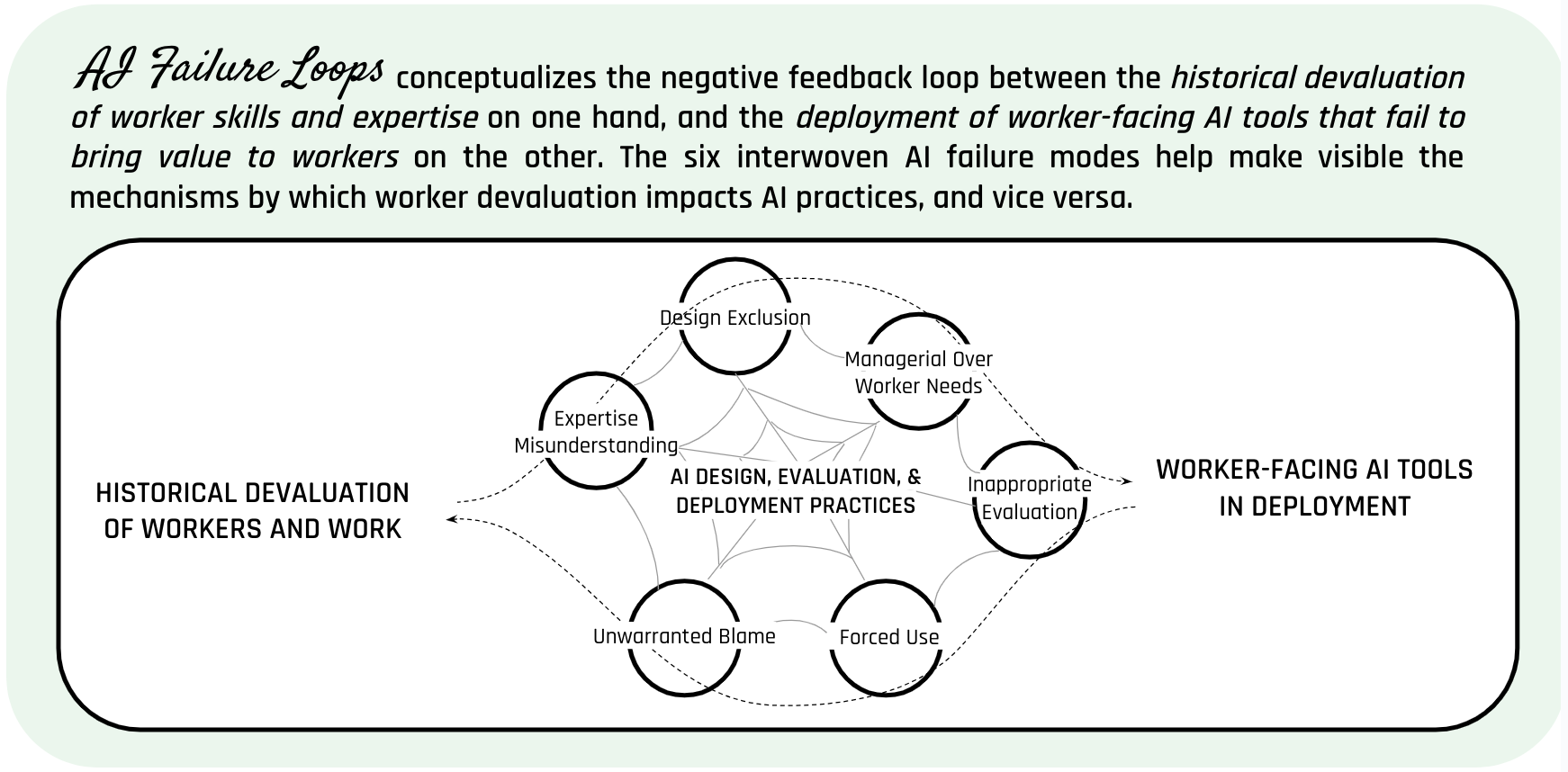}
    \caption{A visual overview of \textit{AI Failure Loops}. \edit{The six failure modes (in circles) that contribute to the dynamics of the \textit{AI Failure Loops} exist within a web to illustrate the inter-connected relationship amongst the failure modes.} }%\ken{minor note: should the caption text in this figure be moved to the actual figure caption (e.g., for accessibility)?}}
    \label{fig:loop}
\end{figure*}

\section{\edit{\textit{AI Failure Loops}}} \label{sec:AIfailureloops}
% Each of our case studies points to the same underlying problem: Society, organizations, and developers misunderstand workers’ expertise, leading to misconceptions about the ability of AI tools to automate or assist workers’ tasks. 
%\edit{AI failures in devalued occupations often point to the same underlying problem: Society, organizations, and developers misunderstand workers’ expertise, leading to misconceptions about the ability of AI tools to automate or assist workers’ tasks. In this section, we formalize how and why this happens through our conceptualization of \textit{AI Failure Loops}.} 
\edit{\textit{AI Failure Loops} arise at the confluence of two dynamics: over-confidence in AI capabilities on one hand, and under-confidence in worker capabilities on the other. We conceptualize \textit{AI Failure Loops} as negative feedback loops that occur when \edit{the devaluation of worker expertise} introduces or reinforces threats to the creation of useful worker-facing AI tools, in turn leading to the deployment of worker-facing AI tools that further amplify harms from \edit{worker} devaluation. Based on prior literature discussed in the Case Studies (Section~\ref{sec:case_studies}), we identify six interwoven AI failure \edit{modes} that shed light into how this can happen: }
\edit{
\begin{itemize}
    \item \textbf{Expertise Misunderstanding}: AI tools in devalued occupations are often motivated by misunderstandings of workers’ expertise. When designing, evaluating, or deploying AI, this may manifest in workers having their expertise questioned. 
    \item \textbf{Managerial Over Worker Needs}: AI tools in devalued occupations are often designed to meet broader organizational goals and targets, at the expense of worker needs. 
    \item \textbf{Design Exclusion}: Workers in devalued occupations tend to be excluded from the design process. If they are involved as one of multiple stakeholders in a design process, their perspectives may be considered a second priority to other stakeholders. 
    \item \textbf{Inappropriate Evaluation}: Evaluations of AI tools in devalued occupations tend to be designed in ways that advantage AI performance and disadvantage workers, further promoting misunderstandings of their expertise and the nature of their work. 
    \item \textbf{Forced Use}: Workers in devalued occupations often do not have the power to refuse technology use.
    \item \textbf{Unwarranted Blame}: Workers in devalued occupations tend to receive unwarranted blame or responsibility for \edit{shortcomings or harms observed from AI systems during} deployment. 
\end{itemize} }

%\jordan{Note: I edited the last paragraph of the failure loops section to change the tense in accordance with the section move.}
\edit{Figure~\ref{fig:loop} provides a high-level visual overview of \textit{AI Failure Loops}. As illustrated by the web, the six failure \edit{modes}
%\footnote{While we identified these six AI failure \edit{modes} based on our analysis of the three case studies and prior literature, this set is not necessarily exhaustive. We hope future work will build upon and elaborate on our account of \textit{AI Failure Loops}. \ken{perhaps delete this footnote? maybe this can be mentioned in the text or saved for the Discussion or Conclusion section} \jordan{+1 to Ken: I also think maybe just a quick sentence or even clause explaining what a "failure mode" is} would suffice} 
are mutually reinforcing, where the presence of one failure mode can amplify the likelihood of others. \edit{As we will see in the Case Studies (Section~\ref{sec:case_studies}), devaluation and misunderstanding of worker expertise are at the core of these dynamics.} In our first case, % in the case studies, we saw how 
we will show how misunderstandings about how social workers make child maltreatment screening decisions (\textit{Expertise Misunderstanding}) led to the use of AI evaluation measures that were fundamentally misaligned with their decision-making goals (\textit{Inappropriate Evaluation}). %We also observed how 
Then, we will describe how home health aides are systematically excluded from the design of new technologies (\textit{Design Exclusion}), in part, because they are made invisible by organizational leaders who did not perceive them to be an integral part of the healthcare system (\textit{Expertise Misunderstanding}). Instead, ``higher status'' healthcare professionals' and institutional needs and goals were prioritized in design efforts (\textit{Managerial Over Worker Needs}). %Finally, in the K-12 teaching case study, 
Finally, we will show how K-12 teachers' exclusion from adoption decisions of new AIED technologies (\textit{Forced Use}) disincentivized AI developers from understanding teachers' work practices, limiting their ability to design AI tools that effectively complement teachers (\textit{Expertise Misunderstanding}). When deployed classroom technologies \edit{fall short of} achieving their envisioned potential, AI developers and school districts often attribute these \edit{shortcomings} to teachers' purported incompetence (\textit{Unwarranted Blame}), with inadequate consideration for potential technology design \edit{issues}.}

\section{Case Studies} \label{sec:case_studies}
%Through these cases, we ground an understanding of the driving factors, properties, and impacts of \textit{AI Failure Loops}, which we synthesize in Section~\ref{sec:AIfailureloops}.
%Organizations have invested in AI innovations intended to assist or automate workers' tasks in feminized labor settings, with hopes to alleviate worker shortages, burnout, and efficiency. However, these technologies are often perceived as missed opportunities to actually support and enhance their work~\citep{holstein2019,saxena2021framework,kawakami2022improving,kawakami2022care,persson2022caregivers}. Our case studies will discuss three such examples of AI deployments in feminized labor contexts: 
\edit{To illustrate the concept of AI Failure Loops, we present three case studies of AI deployments in feminized labor, each emphasizing different failure modes: AI-based risk assessment tools in social work, AI for home healthcare, and AI tutoring systems in K-12 teaching. We provide additional context on each occupation in Table~\ref{table:historical_context}.}

% \jordan{To save words, I think we can delete this preamble because it is already introduced in the last paragraph of the prior section.} -> can remove if we have to cut words for the submission ? 

\begin{table*}[t]
\sffamily
\caption{\redit{Additional historical and social context on the three feminized occupations presented in our case studies (social work, home health, and K-12 teaching in the United States), drawn from existing academic and grey literature.}}
\label{table:historical_context}
  \centering
    \footnotesize
    \setlength{\tabcolsep}{7pt}
    \begin{tabular}{p{0.5\columnwidth}p{0.5\columnwidth}p{0.8\columnwidth}}
    \textbf{Origins} & \textbf{20th Century} & \textbf{Today}\\ 
 \toprule
\rowcolor{mygray}\multicolumn{3}{c}{\textbf{Social Work}}\\
\redit{Founded in 1880s by women organizing settlement house movement, a reformist social movement dedicated to bridging resource gaps between rich and poor families~\citep{socialwork}.} & \redit{Formalized into profession, with requirements for Master's training and state licensing to practice in most U.S. states~\citep{socialwork}.} & \redit{Social workers continue to be predominantly women and disproportionately Black or Hispanic~\citep{socialworknums}. Continued worker devaluation: Paid relatively little compared to other fields requiring similar levels of education~\citep{socialworkunderpaid}; workers voice they feel their expertise is misunderstood and devalued in their workplaces and society~\citep{socialwork}.} \\
\rowcolor{mygray}\multicolumn{3}{c}{\textbf{Home Health}}\\
\redit{Considered part of domestic life for most families in early 19th century, while professional home care was reserved for wealthy families. By late 19th century, women began organizing efforts to send nurses to homes of poorer families~\citep{buhler2003no}.} & \redit{The Fair Labor Standards Act of 1938 outlaws exploitative practices and instituted an 8-hour workday but home health aides are explicitly excluded~\citep{newdeal}.} & \redit{Home health aides continue to be predominantly women and disproportionately Black or Hispanic~\citep{aidelow}. Continued worker devaluation: Majority of workers lack benefits like paid time off, sick leave, or employer-sponsored health insurance~\citep{banerjee2021setting}; majority of workers live below poverty line~\citep{aidelow}; workers often not seen as part of healthcare team, despite being integral to healthcare system~\citep{osterman2018improving}. }\\
\rowcolor{mygray}\multicolumn{3}{c}{\textbf{K12 Teaching}}\\
\redit{Originally conceived as male-dominated occupation that men partook in during farming off-seasons or prior to moving into more well-established professions. Women begin to enter the profession in the late 1880s~\citep{rury1991education}.} & \redit{ Profession is predominantly occupied by women, except among administrators~\citep{teachingWomen}. Profession becomes increasingly formalized with state licensing and university education program requirements~\citep{teacherlicense}.} & \redit{K-12 teachers continue to be predominantly women, except in leadership positions (e.g., principals, superintendents) which continue to be male-dominated~\citep{teachingFem}. Continued worker devaluation: Paid relatively little compared to other fields requiring similar levels of education~\citep{flores2023notions}; workers report high levels of stress and burnout~\citep{steiner2021job}.}\\
\bottomrule
\end{tabular}
\end{table*}

\subsection{AI-Based Risk Assessment Tools Have Amplified Misconceptions of Social Workers' Expertise} \label{social_work}
\subsubsection{\textbf{Context}}

To address widespread staffing and resource shortages, social work agencies across U.S. state and local governments have started to deploy AI-based risk assessment tools in areas like housing allocation for unhoused individuals and screening for child maltreatment cases. %Developers and agency administrators claim these tools help improve the efficiency and accuracy of frontline human decision-making~\citep{kawakamistudying2024,jorgensen2022making}, often in lieu of hiring more social workers. However, subsequent research has found that workers grappling with these AI tools struggle to integrate them into their decision processes~\citep[e.g.,][]{saxena2020human,kawakami2022improving,holten2020,jorgensen2022making}. 
One prominent example of such a tool is the Allegheny Family Screening Tool (AFST), a predictive optimization-based~\citep{wang2022against} tool intended to assist social workers in making screening decisions around potential child maltreatment cases. The model is trained on historical administrative datasets to predict the likelihood of a child being mistreated, using the likelihood a child will be removed from their home within two years as an imperfect proxy. 
Given administrative data on a particular family, the AFST outputs predictions in the form of a risk score between 1 and 20. This class of AI-based risk assessment tool has been adopted across other agencies in the U.S., New Zealand, and Denmark, and applied to other social work decision-making tasks (e.g., foster care placements, temporary lodging placements). In initial years of deployment, the social work agency received both widespread praise and criticism for its deployment of the AFST. On the one hand, the agency was lauded for publishing public-facing evaluations on the ethical implications and technical performance of the model (e.g.,~\citep{AFSTdocumentation}). On the other hand, academics, advocates, and journalists heavily criticized the agency for deploying an AI model with biased performance against Black and disabled communities~\citep{afstjustice,gerchick2023devil} and poor construct validity~\citep{coston2022validity}, with some calling to end the use of such tools.

\subsubsection{\textbf{AI Failures in Social Work}}
\redit{We turn our attention to how the kinds of AI design and deployment challenges discussed in these prior critiques} may have been influenced by how workers in this domain are perceived and treated by agency leadership and AI developers. This section revisits prior literature on the design and use of the AFST, using \textit{AI Failure Loop} as a lens. 

\paragraph{\textbf{AI tool design fundamentally misunderstands social workers' expertise and overlooks their needs} }
    
The social work profession has a ``glass ceiling,’’ where workers have minimal opportunities to advance into organizational decision-making roles. As a consequence, there are often large silos between agency leaders and social workers, so those making decisions impacting social workers may have limited procedural understanding of the expertise and on-the-ground knowledge social work requires~\citep{holten2020,kawakami2022improving,saxena2021framework}. 

In the AFST case, the agency issued an open call to create a new tool to support child maltreatment call screeners in making more efficient decisions. Despite  social workers being the main users of the tool, the tool creators did not treat them as stakeholders in the design process. For example, social workers' perspectives on the AI tool were collected only after the model was already developed, using narrowly focused surveys confirming their comprehension of a brief training on the tool, while limiting broader feedback on the tool's overall problem formulation and design~\citep{kawakami2022improving}.

As a result, both researchers and workers have since pointed to fundamental flaws in the AFST's design, which may have been avoided had workers been meaningfully involved in the design process~\citep{gerchick2023devil,kawakami2022improving}. For example, the longer-term outcomes this AI tool was trained to predict (risk of out-of-home placement over the next two years) fundamentally conflicted with how workers were trained and legally-required to make decisions (i.e., based on assessments of immediate risks to the child). Therefore, once deployed, workers expressed concerns about the tool's validity and uncertainty around how they were  expected to interpret and use its outputs~\citep{cheng2022child,kawakami2022improving,kawakami2022care}. 

More broadly, workers questioned the validity and usefulness of the tool's overall formulation: statistical risk prediction based on government administrative records. Instead of presenting opaque predictions derived from the shallow information available in administrative data, workers believed it would have been more helpful to have AI support \textit{them} in making sense of available evidence---leveraging their on-the-ground knowledge and qualitative understanding of case-specific context as human experts~\citep{kawakami2022care}.

\paragraph{\textbf{Developers interpret worker complaints around AI as signals of worker deficits}}

As illustrated above, social workers are often excluded from pre-deployment design and deployment decisions of AI tools that intend to support their work~\citep{kawakamistudying2024}. When workers are given opportunities to provide feedback on AI tools post-deployment, their complaints are often dismissed as an inability to understand a complex technology. For example, in the AFST case, one worker described the types of responses they would receive after submitting feedback forms on the AI tool’s risk score as follows: 
\begin{quote}
``The input does not feel like a two-way street. [We] are told why we’re wrong, and [that] we just don’t actually understand algorithms versus, like, maybe you [the workers] are observing something [our algorithm] missed, you know? [...] In the past, it’s felt like when we’ve taken advantage of that function [to give feedback on the AFST score], it has kind of resulted in just explaining to you why the score is actually right’’~\citep{kawakami2022improving}. 
\end{quote}

\paragraph{\textbf{Social workers are forced to use AI, even when they disagree}} \label{sw_force}
Social workers cannot simply refuse technology use~\citep{kawakami2022improving,kawakamistudying2024,saxena2021framework}. Organizational policies govern how often they can apply discretion with the  model. For example, social workers were required to use the AFST and were expected not to disagree ``too often.'' Workers feared that violating this expectation could jeopardize their employment, so they felt pressure to agree with the AI model, often against their best judgment~\citep{kawakami2022improving}. Nonetheless, workers still exercised their ability to disagree with the AI on a ``budget,'' to try to do right by families under these constraints~\citep{cheng2022child}.

Similar AI-based risk assessment tools have been deployed in criminal justice. For example, COMPAS assists judges in making recidivism risk assessments~\citep{dressel2018accuracy}. Despite some similarities between the risk assessment tools in social work versus criminal justice (e.g., excluding workers from the design process), we argue there are some notable differences. Given judges' relatively higher-status and organizational power, they have greater ability to refuse technology use by ``ghosting the machine’’~\citep{pruss2023ghosting}, ignoring risk scores at their discretion or discarding the entire risk assessment model, without fear of employment consequences. 

\subsection{``AI for Healthcare'' Has Left Behind Devalued Healthcare Experts}\label{healthcare}
\subsubsection{\textbf{Context}}
Governments and private investors have thrown billions of dollars into healthcare AI innovations---and this number is only increasing each year~\citep{esposito2016supporting,yang2019unremarkable,yildirim2023creating}. Many of these innovations target home healthcare. Yet, today, there exist few, if any, technical innovations that actually \textit{support} home health aides~\citep{kuo2022understanding}---workers, the majority of whom are women and people of color, who support elderly people and people with disabilities in everyday personal care  (e.g., bathing, washing, monitoring health challenges) and household responsibilities (e.g., cooking, cleaning, shopping).  \redit{Instead, existing technologies in home health have tended to \textit{automate} worker skills (e.g., care robots~\citep{esposito2016supporting}), \textit{manage} worker practices (e.g., shift matching~\citep{solano2025running}), or provide support to other stakeholders like patients (e.g., interactive voice assistants~\citep{bartle2022second}).}  
For example, to help address the immense care work shortage, several countries--across Europe, North America, and East Asia--have heavily invested in exploring the potential of care robots~\citep{esposito2016supporting}. In fact, care robots for home healthcare applications are among the main investments in robotics applications~\citep{esposito2016supporting}. However, a growing body of work has recognized that, if not designed carefully, these systems may disempower workers and amplify challenges they face already~\citep{fox2019managerial,tseng2020we,persson2022caregivers}. \redit{These tendencies to decenter home health aides in AI innovation efforts---in home health and healthcare more broadly---}has occurred amid increasing acknowledgement that home health aides are the cornerstone to systems of care, that demands for their labor are increasing, and that institutional support for their wellbeing and labor has remained stagnant~\citep{aidelow}. 

%In this case study, we critically examine technology deployments that intend to support home health aides, particularly to understand how home health aides' persisting devaluation may help explain failures of AI technologies to genuinely support them.

\subsubsection{\textbf{AI Failures in Healthcare}}
This failure to deploy technical innovations that support home health aides can be understood as a failure of existing efforts to consider the impacts of and on the devaluation of home health aides.
This section revisits \textit{why } research and design communities have failed to design AI technologies that genuinely support home health aides, borrowing \textit{AI Failure Loop} as a lens. 

\paragraph{\textbf{Home health aides are rarely designed for}}
In the past decade, ``AI for healthcare'' has grown substantially as both an area of academic research and capitalist speculation. These efforts often center ways to better meet \textit{doctors’} needs. However, nurses and home health aides—who also play critical roles in healthcare—are rarely designed for or with~\citep{kuo2022understanding}. 

While healthcare services rely on collaboration among various professions--including home health aides, nurses, and doctors--most existing efforts to deploy AI technologies overlook this. Instead, the majority of AI for healthcare efforts in the past decades focused on supporting doctors (e.g., physicians, radiologists, surgeons), for example, through new clinical decision support tools~\citep{yang2016investigating,yang2019unremarkable,sendak2020real}. We argue that the decentering of home health aides in ``AI for healthcare’’ efforts has happened, in part, due to misconceptions of the nature of their labor and organizational dynamics that further entrenched these misconceptions.

While home health aides provide many of the same services as nurses in healthcare institutions~\citep{hittle2016complexity}, their labor is performed within their clients' homes rather than in institutions. Prior literature has described this physical isolation from healthcare institutions has made their labor less visible to other healthcare professionals and policymakers, and created associations with informal home care work~\citep{cho2023care}, challenging the profession's perceived legitimacy along the way. 

These misconceptions of home health aides’ expertise, and along with it, their systemic devaluation, are also observed through how home health aides are treated within healthcare institutions. In \textit{Improving Long-Term Care by Finally Respecting Home-Care Aides}~\citep{osterman2018improving}, Osterman recalls a conversation with a nurse: ``no one pays attention to home-care aides . . . . When something changes they notice but are not taken seriously.’’ These misjudgements may not be explicitly acknowledged nor understood. For example, Osterman recalls conversations with other healthcare professionals and stakeholders, observing:

\begin{quote}

``other medical professionals, home care agency executives, directors of nursing homes, insurance company executives, and state and federal policy makers, I consistently heard comments that began with a bow to what wonderful people direct care workers are, and then went on to demean their skills and potential’’~\citep{osterman2019improving}

\end{quote}

Even as decision-makers in healthcare (e.g., doctors, hospitals, state regulators, insurance companies) verbally acknowledge home care aides, they ``typically do not think that home care aides can be real members of a care team’’ ~\citep{osterman2019improving}. Osterman describes this happens, in part, because home care aides' skills are overlooked, instead misunderstood as that of ``unskilled companions [and] glorified babysitters’’~\citep{osterman2019improving}. \redit{Ming et. al. observes that home care aides recognize these misjudgements: ``Some people see us as a maid [...] We’re not just here to clean your house [..] Maid are different from an aide''~\citep{ming2023go}. \cite{ming2023go} further cautions that failing to account for the invisibility of aides' work may contribute to the deployment of AI systems that enforce compliance to impoverished notions of ``care''---which, in turn, may further devalue and make invisible home health aides' work. }

In fact, most existing technologies intending to support home care agencies have geared towards supporting higher-status workers in home health agencies (e.g., doctors, nurses), rather than directly helping home health aides. A literature survey from 2022 found only nine studies contributed technologies to support home health aides~\citep{kuo2022understanding}. Of these, the majority (7/9 studies) were technologies where home health aides were only one of several potential users. The authors found no studies with a long-term evaluation on how the technology impacted home health workers. 

\paragraph{\textbf{Existing technology ``for’’ home health aides have failed to emulate or enhance their work}}
Researchers and policymakers have long looked for ways to sustain the growing population of elderly people---much of which is currently supported through home health aides~\citep{bedaf2015overview,hendry2004care}. While there has been limited research on technologies to support home health aides, there is a comparatively large number of papers exploring ways to automate home-based tasks needed by elderly people, including tasks home health aides usually complete~\citep{esposito2016supporting}. Many of these technical deployments have come in the form of care robots. However, despite the massive amounts of funding dedicated towards advancing the design of care robots, actual deployments of care robots have often been deemed failures~\citep[e.g.,][]{persson2022caregivers,sharkey2014paro}. Most of this literature has discussed failures with respect to meeting elderly people’s care needs.% In this section, we discuss how care robots have also failed to bring value to home health aides themselves.

While care robots may have been intended to help alleviate the increasing demand for home care workers, the few studies that account for impacts on care workers have observed impacts in the opposite direction. 
For example, in a scoping review of 27 papers on the deployment of care robots, researchers concluded the effectiveness of care robots is ``mixed’’ at best. It further observed how robots can instead \textit{increase} caregivers’ workload: care robots ``seldom work as a shortcut to increased efficiency or time effectiveness'' as workers must undergo the labor of maintaining and operating the robot~\citep{persson2022caregivers}. Another study found care workers experienced emotional demands, grappled with new ethical challenges, and felt decreased job satisfaction when care robots were deployed in their workplace~\citep{wright2018tactile,wright2019robots}.

\redit{Beyond care robots, \cite{solano2025running} discuss similar challenges with other forms of home health technologies, like AI-based shift management technologies that match home health aides with patients. Home health aides reported feeling concerned about how matching algorithms fail to account for the \textit{human} dimensions of their work, like personality, compassion, and empathy, that impact the quality of matching decisions~\citep{solano2025running}. Others described how AI-based shift management technologies may further increase the invisibility of home health aides' work by taking away the already minimal amount of contact they have with home health agencies~\citep{solano2025running}.} 

\paragraph{\textbf{Institutional power imbalances incentivize involvement of higher-status workers in AI design}} \label{health_power}
Home health aides are rarely consulted during the design process, even for technologies  directly impacting them~\citep{kuo2022understanding}. This creates a stark contrast with the development of technologies that support doctors. In this section, we briefly discuss how having organizational power as a ``high status’’ worker can create incentives to avoid the \textit{AI Failure Loop}. In doing so, we shed light onto the power and labor dynamics at play in low-status occupations, that lead to deviations in how they are impacted by AI deployments. 

The HCI, ML, and medical research and development communities 
have long been interested in innovating new AI tools to support medical decision-making. Even with decades of attempts, successfully deploying AI tools into real-world healthcare contexts has remained a challenging task~\citep{yildirim2023creating}. Once deployed, the majority of AI innovations have failed to bring value to doctors~\citep{yang2016investigating}. Much like in social work, early versions of decision support tools were often created in isolation from the professional decision-makers themselves: Medical experts were not consulted during the design process. Academics and hospitals found that, even after spending millions of dollars and decades of research innovating on medical decision support tools, doctors were simply ignoring them once they were deployed in practice~\citep{yang2019unremarkable}. This ability to ignore the AI tool--to have full agency over decisions to ignore or disagree with an AI tool--incentivized researchers and developers to consult doctors in future AI innovation attempts. 

We argue that doctors' refusal to use AI innovations incentivized researchers and developers to more closely examine \textit{why} doctors were not adopting them~\citep[e.g.,][]{wyatt1995commentary,yang2019unremarkable}. To ensure they were creating AI tools doctors would actually use once deployed, the academic and developer communities may have been more incentivized to account for doctors' needs and desires (as evidenced by the gradual surge of academic papers documenting medical experts' needs~\citep[e.g.,][]{yang2016investigating,cai2019hello} and involvement in AI creation ~\citep[e.g.,][]{yang2019unremarkable,sendak2020real,yang2016investigating}). 

In comparison, home health aides, like social workers, are not as well-respected in the healthcare system, making it less likely for them to have agency over technology deployments like the ability to refuse technology use. 
Moreover, because post-deployment evaluations of home health aides' perceptions of AI tools are comparatively scarce~\citep[e.g.,][]{kuo2022understanding}, their negative impacts on workers may also be more likely to go unnoticed.

\subsection{Persistent Visions of Automation in Education Have Further Devalued Teachers' Expertise}\label{k12teaching}
\subsubsection{\textbf{Context}}
%K-12 teaching in the United States was originally conceived as a male-dominated occupation, %in the colonial period, 
%but gradually became dominated by women from the late 1800s as men started to move into more lucrative occupations. From the start, women teaching K-12 education struggled to establish themselves as working professionals. This was, in part, because when teaching first formalized as a profession, it was treated as a transient activity that men partook in during off-seasons for farming or prior to moving onto more well-established professions~\citep{rury1991education}. Even as U.S. society began to value K-12 education as an important social good, respect towards women teachers still staggered. Instead, male administrators became an increasingly authoritative figure supervising female teaching~\citep{teachingWomen}. 

%Today, teaching for K-12 education in the United States is still commonly associated with women~\citep{miller2000women}. Echoing the ``glass ceiling'' felt in other feminized occupations, the percentage of women in K-12 teaching starkly decreases as the pay and status associated with the position increases. For example, while the majority of K-12 teachers are women, men continue to dominate half the population of principals and three-fourth of district superintendents~\citep{teachingFem}.  

Researchers and developers have long explored ways to introduce new technologies into K-12 education, with hopes to transform the schooling experience. The vision behind AI in education (AIED) research %AI in Education (AIED) research has often been motivated, in part, by the goal of decreasing divides between privileged and underprivileged learners~\citep{holmes2023unintended}. The vision 
is that every student will be able to work at their own pace and get the personalized support they need~\citep{holmes2023unintended,edhistfuture}. 
However, these technologies have often been designed with a focus on supporting student learning, without considering the needs and practical constraints of the \textit{teachers} who are asked to use them in their classrooms ~\citep{baker2016stupid,holmes2023unintended,holstein2017intelligent,roll2016evolution}.  
%In this section, we use the \textit{AI Failure Loop} as a lens to discuss prior literature on AIED design and evaluation \edit{practices and challenges} that have contributed to these trends. 

%\jordan{The paragraph above makes a lot of claims without citation.}

\subsubsection{\textbf{AI Failures in K-12 Teaching}}
As in the prior two case studies, we see that AI deployments in education have been designed and evaluated at the confluence of two persistent dynamics: policymakers, companies, and school districts have long lauded the 
potential of AIED technologies to radically transform K-12 education, while, at the same time, these groups have devalued and misunderstood K-12 teachers’ expertise. 
In this section, we examine how AI design, evaluation, and deployment practices have been impacted by these dynamics.

\paragraph{\textbf{Designing to help students learn, at the expense of helping teachers teach}}
There is a long history of AI development in education guided by the ultimate vision of automating away the role of human teachers~\citep{edhistfuture,edtechkey}. 
Over the past forty years, research and practice in AIED has largely centered students’ needs, often without explicit consideration for teachers’ needs and constraints in the classroom~\citep{baker2016stupid,holmes2023unintended,holstein2017intelligent,roll2016evolution}.
Holmes refers to these deployments as ``student-focused AIED’’---AI technologies aiming to automate one or more of teachers’ tasks for the stated goal of helping students learn, for example, including tutoring systems, virtual writing assistants, automatic writing evaluation, and chatbots. By contrast, Holmes finds that ``teacher-focused AIED’’ has remained largely speculative, confined to academic research prototypes rather than commercial systems~\citep{holmes2023unintended}. 
Beyond the research community, commercialization practices help explain the prevalence of student-focused AIED technologies at the expense of teacher-focused AIED. Companies sell AIED technologies to schools and districts, and past research has found that teachers may be given little say in whether or not to adopt a given technology~\citep{holstein2017intelligent,utterberg2021intelligent}. 

By reducing K-12 teachers to mere service providers responsible for operating technology and maintaining the classroom~\citep{holmes2023unintended}, many commercialized AIED deployments require teachers to ``accommodate’’ to the technology. Indeed, prior work examining AI deployments observed how teachers often struggle to make AIED technologies work in their classroom, without having to adjust their practices in unreasonable ways or take on new responsibilities along the way~\citep{enyedy2014personalized,utterberg2021intelligent}. 
Observing math teachers using an AI tutoring systems in their classrooms, 
Holstein et al. and Utterberg Moden et al. found that teachers were forced to grapple with numerous contradictions designed into the AIED system, eventually causing them to abandon the system. 
For instance, while the AI tutoring system aimed to help students work at their own pace, this ran counter to pedagogical practices around collaborative learning and discussion: Teachers had a responsibility to maintain a cohesive classroom environment, where students were roughly synchronized in the content they were learning.%, but the tutoring system failed to accommodate this need 
~\citep{holstein2017intelligent,utterberg2021intelligent}. 

Moreover, these failures \edit{to support teachers} may be challenging to capture post-deployment because existing AI evaluation measures focus on improving student learning and achievement outcomes with little regard for teacher experiences. As was the case with AI deployments for social workers and home health aides, the emotional and physical labor involved in integrating and operating AI technologies into the workplace are not considered in assessments of new AI deployments. 

\paragraph{\textbf{``Teacher bashing’’: Teachers receive unwarranted blame for failed AIED deployments }}\label{teacher_bashing}

Despite the labor and skill required to make AIED technologies work in their classroom, teachers are disproportionately blamed %or held responsible 
for \edit{limited adoption and use of AIED technologies in classrooms}%failures
~\citep{holmes2023unintended,edtechkey,edhistfuture}. 
Indeed, there is often an assumption that the reason AIED tools are not well-implemented is because teachers are too disinterested or unskilled to use the technology effectively. In their book, ``Education and Technology: Key Issues and Debates,’’ Selwyn reflects: 

\begin{quote}
``As is often the case with debates over the `failures' of education systems, `blame' for the restricted use of technology in schools, colleges, and universities has tended to be attributed most readily to the perceived shortcomings of teachers [...] teachers have been deemed to be too old, disinterested, or incompetent to integrate digital technology [...] It is all too easy for enthusiastic academic commentators to indulge in `teacher bashing' and portray teachers as outmoded, obstructive, or ignorant.''
~\citep{edtechkey}
\end{quote}

\paragraph{\textbf{Limited incentive to change AIED development practices}}
Prior work has documented instances in which teachers were able to abandon AIED deployments that failed to bring them value~\citep[e.g.,][]{holstein2017intelligent,utterberg2021intelligent}. However, even when teachers unite to refuse technology use, they are not guaranteed an escape from \textit{AI Failure Loops} longer term. 
As discussed in Section~\ref{sw_force} (judges) and Section~\ref{health_power} (doctors), higher-status workers may be able to simply ignore the outputs of an AI tool, creating natural incentives for future AI developers to account for their needs and desires in future AI creation attempts. 
On the other hand, lower status workers like teachers may manage to escape from an instance of a failed AI deployment, but these escapes alone are unlikely to prompt meaningful improvements in AI development practices.
As discussed above, school districts and developers may still disproportionately blame teachers for %attribute blame for 
\edit{low adoption and use of AIED technologies} %to teachers, 
missing opportunities to identify ways to design AI technologies that genuinely support teachers. 

\section{Discussion: Toward Pro-Worker AI}

% In this section, we draw further attention to the case of feminized labor, discussing how feminized labor occupations--as a particular form of devalued labor--uniquely intersects with \textit{AI Failure Loops} (Section~\ref{femme_case}). We then situate our core arguments within prior literature theorizing \edit{reasons for} workplace AI failures, to discuss the extent to which the dynamics and challenges presented in this paper are driven by recent AI developments versus recurring patterns from history (Section~\ref{whats_new}). Looking forward, we present four critical questions towards a \textit{pro-worker} future for AI and society. 

\edit{In this section, we discuss implications for researchers and practitioners to support the study and development of pro-worker AI. We begin with three critical questions (Section~\ref{q1}--~\ref{q3}) to guide future research, policy, and practice towards a pro-worker future for AI and society. % then end with a discussion on unique considerations for the case of feminized labor (Section~\ref{femme_case}).  
\redit{Each question raises implications to help \textit{speak back against} currents that devalue worker expertise through AI practices. Drawing on lessons from our case studies, we explore how \textit{AI Failure Loops} might be avoided by attending to the ways design processes account for power dynamics (Subsection~\ref{q1}), the ways measurement can be used to make misleading claims about human versus AI performance (Subsection~\ref{q2}), and opportunities to bridge the gaps between pro-worker AI in theory and in practice (Subsection~\ref{q3}).} }
We end with a discussion of unique considerations for the case of feminized labor (Section~\ref{femme_case}).

% \subsection{Toward Pro-Worker AI}

% \subsection{Whose expertise is considered valuable?}
% \subsection{Value Worker Expertise}
\subsection{(How) can design processes manipulate power?}\label{q1}
\edit{Our findings demonstrate how the design of workplace AI can reproduce and exacerbate occupational devaluation. The everyday work of devalued occupations, including but not limited to feminized labor, is often obscured from those in comparatively ``high status'' positions in the same organizations~\citep{blomberg1996reflections}. Yet, across our cases, AI developers routinely privileged the knowledges of ``experts'' to \textit{speak for} devalued workers. Rather than involving social workers themselves, developers consulted social work academics (Section~\ref{social_work}); rather than involving teachers, developers consulted school district administrators (Section~\ref{k12teaching}). In doing so, AI developers further perpetuated the invisibilization of feminized labor and the devaluation of workers' expertise~\citep{haraway2013situated}. In failing to include — let alone center — social workers and teachers in the design process, developers created tools that did not work for these frontline workers.}

\edit{Centering historically devalued workers in the design and research of AI systems poses unique challenges. Echoing longstanding concerns regarding gatekeepers in participatory design \citep{spiel2020details}, relying on managers to access workers may exert undue influence on research and design decisions. Instead of expecting leaders to open a pathway to involve workers, we encourage researchers and practitioners to buy workers’ time and form collaborations with existing labor unions and worker advocates~\citep{spektor2023charting}. Researchers should also critically examine worker participation in AI development. Workers may feel forced to participate \citep{suchman2011anthropological} or be included \citep{hoffmann2021terms} in AI design processes with which they disagree but cannot reject \citep{delgado2023participatory}. If devalued workers are involved as one of multiple stakeholders, treating their voice as “equal” to other ``high status'' stakeholders may reproduce workplace power imbalances~\citep{fox2019managerial}. Finally, we caution that managers and developers may recuperate the language of ``participation'' by using minor worker involvement to justify AI deployments, also known as \textit{participation washing}~\citep{sloane2022participation}.}

\subsection{(Why) do we need to measure human performance against AI?} \label{q2}% measure what counts for workers 
\edit{The measurements used to evaluate AI models are but one of many representations of the world \citep{boyd2012critical}. Our case studies demonstrate how the design of workplace AI evaluation metrics can further marginalize devalued workers. As we described in Section~\ref{social_work}, managers measured worker versus AI performance according to AI developers' flawed predictive target. Even though this predictive target misaligned with workers' underlying decision-making goals~\citep{cheng2022child}, this comparative analysis was used to justify the AI tools' deployment. Moreover, once AI tools were deployed, these flawed measurements were used to track and surveil workers' compliance with AI outputs~\citep{kawakami2022improving}.}

\edit{Our case studies also strongly suggest that we should question the value of ``measuring'' workers in the first place. As noted by CSCW scholars, the question of what work to make (in)visible through datafication can be politically fraught \citep{star1999layers}. Researchers and developers may find that some work cannot or should not be quantifiable, leading to the development of alternative evaluation standards (e.g., worker satisfaction). However, even in occupations where there has been decades of research and development, as is the case with AIED technologies in K-12 education, the vast majority of evaluations for AI deployments have focused on understanding impacts on other stakeholders (e.g., students), without consideration for frontline workers (e.g., teachers') experiences and perspectives~\citep{holmes2023unintended}.} 

\redit{In contrast to current workplace measurement approaches, future research could explore the design of new approaches \textit{with} workers that help make visible the (currently invisible) impacts of AI deployments on their labor. In these efforts, ensuring worker ownership over data collection and disclosure decisions is critical to ensuring that such measurements can actually enable workers to advocate for better technologies and working conditions (or the abandonment of technologies), rather than further surveil~\citep{ming2023go}.}

\subsection{What happens after worker-centered design?} \label{q3}
%\anna{this is previously: What happens after worker-centered design?}
\edit{Our case studies also demonstrate a tension between the theory and practice of pro-worker AI research and design. As illustrated in the education case study (Section~\ref{k12teaching}), institutions may be less incentivized to adopt or commercialize AI technologies that center devalued workers’ needs. This has led teacher-centered AIED tools to remain largely ``speculative''~\citep{holmes2023unintended}, where they exist as research prototypes but not as actual tools used in the classroom. Worker-centered AI tools that do get adopted by real-world organizations may still inadvertently reinforce worker devaluation through organizations’ governance practices. As \textit{AI Failure Loops} explains, workers in devalued positions tend to receive unwarranted blame for unanticipated issues and harms that arise from AI post-deployment, even if they have minimal power to prevent them. If AI harms are only visible to workers, they may also not have power to refuse to use the technology. Even in efforts to deploy and, when appropriate, commercialize AI technologies, researchers and developers must continue to remain wary of---and actively correct for---the impacts of occupational devaluation. }

\edit{Researchers can also support pro-worker AI through policy advocacy aimed at empowering workers to govern AI use and maintenance. For example, future work could design new approaches to involve workers in the creation of organizational policies that govern technology use---including to empower worker involvement in decisions around whether and when to decommission a technology. Future work could also explore ways to expose policymakers to the (often overlooked) impacts of technology on workers, to inform the creation of sector-specific worker-protecting regulations (e.g.,~\citep{zhang2024data}). Finally, future work should continue to formalize mechanisms for worker collective action by, for example, designing new technologies that allow workers to organize around and share experiences with technology use (e.g.,~\citep{irani2013turkopticon}). }

\edit{Finally, our case studies demonstrate the importance of critical research elevating workers' voices and questioning the design of workplace AI systems. For example, the AFST research we described in Section~\ref{social_work} partially contributed to some US states stopping the use of similar child welfare screening algorithms in other US states \citep{june2022oregon}. Likewise, the development of a child welfare risk assessment tool in Denmark was canceled due in part to public scrutiny fueled by critical scholarship \citep{ratner2023algorithmic}. While researchers may not always have the power to ensure that workers are centered in the design of workplace AI systems, }%; hence, the tendency for worker-centered AI research to be more speculative (Section ??). 
\edit{they can play a crucial role in contesting the proliferation of workplace AI systems that reinforce worker devaluation.}

\subsection{The Case of Feminized Labor} \label{femme_case}
\edit{Redirecting toward a pro-worker future for AI may involve some unique considerations in the context of \textit{feminized labor}. For instance, many feminized occupations center on emotional and care-oriented labor that may be difficult or impossible to quantify, increasing risks of misquantifications in AI development. Moreover, workers in service-oriented positions (i.e., most feminized positions) are often trained to decenter themselves, becoming a ``non-person'' in their workplace and to employers~\citep{forsythe1993construction,ming2023go}. Even if AI developers manage to form genuine design partnerships with workers, they may not---without explicit encouragement---advocate for designing technologies that meet their own needs. In past efforts to design technologies \textit{with} feminized workers (e.g., home health aides, K-12 teachers), these workers often centered clients' needs and de-centered their own needs by default during design engagements, unless nudged by designers to do so~\citep{holstein2019,bartle2022second}. 
We anticipate that, with time, redirecting this trajectory will only increase in importance for AI in feminized occupations. When one considers that many of the most devalued occupations are dominated by women, the majority of whom are women of color, it becomes necessary to recognize the presence and prevalence of \textit{AI Failure Loops} as a racial, gender, and social justice issue.}

\section{Conclusion}
\edit{Visions for ``transformative’’ workplace AI technologies that undermine workers are not new. Technologists have long envisioned that AI technologies would ``fix’’ staff shortages in feminized occupations.  The challenges described in this paper—the development of workplace technology without a grounded understanding of workers and their work—are reminiscent of observations made by Forsythe nearly three decades ago~\citep{forsythe1993construction}. Forsythe found that AI developers building expert systems held impoverished understandings of ``work,’’ disregarding social and maintenance activities. Thus, these expert systems failed to model experts’ real-world work practices. Despite the shift from rule-based to machine learning systems, many of Forsythe's observations persist today. }

\edit{Since Forsythe's writing, the scope and breadth of worker tasks that AI developers target have substantially expanded, from systems attempting to provide emotional support for K-12 students to predicting whether a child is at risk of future abuse or neglect. As this trend continues, our case studies provide a cautionary tale: historically devalued occupations are particularly vulnerable to \textit{AI Failure Loops}. We expect recent increases in both AI hype and actual capabilities will exacerbate the over-estimations of what AI can do and under-estimations of workers' expertise.}

\edit{Critically, nothing about the future of AI and work is inevitable. To overcome \textit{AI Failure Loops}, we must learn from past developments that perpetuated the devaluation of feminized labor and contest the societal and organizational devaluation of certain work. Re-imagining AI development practices to empower devalued workers may even help foster such cultural shifts. Just as \textit{AI Failure Loops} are self-reinforced through negative feedback loops, reversing these trends may cultivate \textit{positive} feedback loops between pro-worker AI practices and worker valuation. %Redirecting towards a \textit{pro-worker} future for AI and society requires not just \textit{recognizing} these currents but \textit{speaking back against them}. 
} \edit{This paper extends discussions of how researchers, policymakers and practitioners can help redirect towards a \textit{pro-worker} future for AI and society, by drawing lessons learned from how worker devaluation has shaped AI practices, and vice versa.}

\begin{acks}
We would like to thank the reviewers for their excellent reviews that helped strengthen our paper. We are also grateful for the thoughtful feedback shared by Alicia DeVrio, Jodi Forlizzi, Nari Johnson, Joy Ming, Tina Park, and David Widder.
\end{acks}

\bibliographystyle{sageh}

\begin{thebibliography}{99}
\providecommand{\natexlab}[1]{#1}
\providecommand{\url}[1]{\texttt{#1}}
\providecommand{\urlprefix}{URL }
\expandafter\ifx\csname urlstyle\endcsname\relax
  \providecommand{\doi}[1]{DOI:\discretionary{}{}{}#1}\else
  \providecommand{\doi}{DOI:\discretionary{}{}{}\begingroup \urlstyle{rm}\Url}\fi

\bibitem[{tea(2018)}]{teacherlicense}
 (2018) Teacher education: From revolution to evolution.
\newblock \urlprefix\url{https://teachereducation.steinhardt.nyu.edu/teacher-training-evolution/}.

\bibitem[{per(2022)}]{persson2022caregivers}
 (2022) Caregivers’ use of robots and their effect on work environment--a scoping review.
\newblock \emph{Journal of technology in human services} 40(3): 251--277.

\bibitem[{soc(2022)}]{socialworkunderpaid}
 (2022) School of social work study confirms human services workers are underpaid.
\newblock \urlprefix\url{https://socialwork.uw.edu/news/school-social-work-study-confirms-human-services-workers-are-underpaid/}.

\bibitem[{Acemoglu et~al.(2023)Acemoglu, Autor and Johnson}]{acemoglu2023can}
Acemoglu D, Autor D and Johnson S (2023) Can we have pro-worker {AI}?
\newblock \emph{CEPR} .

\bibitem[{{AP News}(2023)}]{afstjustice}
{AP News} (2023) Child welfare algorithm faces justice department scrutiny.
\newblock \urlprefix\url{https://apnews.com/article/justice-scrutinizes-pittsburgh-child-welfare-ai-tool-4f61f45bfc3245fd2556e886c2da988b}.
\newblock Online.

\bibitem[{Baker(2016)}]{baker2016stupid}
Baker RS (2016) Stupid tutoring systems, intelligent humans.
\newblock \emph{International journal of artificial intelligence in education} 26(2): 600--614.

\bibitem[{Banerjee et~al.(2025)Banerjee, Gould and Sawo}]{banerjee2021setting}
Banerjee A, Gould E and Sawo M (2025) Setting higher wages for child care and home health care workers is long overdue.
\newblock \urlprefix\url{https://www.epi.org/publication/higher-wages-for-child-care-and-home-health-care-workers/}.

\bibitem[{Bartle et~al.(2022)Bartle, Lyu, El~Shabazz-Thompson, Oh, Chen, Chang, Holstein and Dell}]{bartle2022second}
Bartle V, Lyu J, El~Shabazz-Thompson F, Oh Y, Chen AA, Chang YJ, Holstein K and Dell N (2022) “{A} second voice”: Investigating opportunities and challenges for interactive voice assistants to support home health aides.
\newblock In: \emph{Proceedings of the 2022 CHI Conference on Human Factors in Computing Systems}. pp. 1--17.

\bibitem[{Bedaf et~al.(2015)Bedaf, Gelderblom and De~Witte}]{bedaf2015overview}
Bedaf S, Gelderblom GJ and De~Witte L (2015) Overview and categorization of robots supporting independent living of elderly people: What activities do they support and how far have they developed.
\newblock \emph{Assistive Technology} 27(2): 88--100.

\bibitem[{Blomberg et~al.(1996)Blomberg, Suchman and Trigg}]{blomberg1996reflections}
Blomberg J, Suchman L and Trigg RH (1996) Reflections on a work-oriented design project.
\newblock \emph{Human--Computer Interaction} 11(3): 237--265.

\bibitem[{Boyd and Crawford(2012)}]{boyd2012critical}
Boyd D and Crawford K (2012) Critical questions for big data: Provocations for a cultural, technological, and scholarly phenomenon.
\newblock \emph{Information, communication \& society} .

\bibitem[{Brynjolfsson et~al.(2025)Brynjolfsson, Li and Raymond}]{brynjolfsson2025generative}
Brynjolfsson E, Li D and Raymond L (2025) Generative {AI} at work.
\newblock \emph{The Quarterly Journal of Economics} .

\bibitem[{Buhler-Wilkerson(2003)}]{buhler2003no}
Buhler-Wilkerson K (2003) \emph{No place like home: {A} history of nursing and home care in the United States}.
\newblock JHU Press.

\bibitem[{Cai et~al.(2019)Cai, Winter, Steiner, Wilcox and Terry}]{cai2019hello}
Cai CJ, Winter S, Steiner D, Wilcox L and Terry M (2019) "{Hello} {AI}": Uncovering the onboarding needs of medical practitioners for human-{AI} collaborative decision-making.
\newblock \emph{Proceedings of the ACM on Human-computer Interaction} 3(CSCW): 1--24.

\bibitem[{Cheng et~al.(2022)Cheng, Stapleton, Kawakami, Sivaraman, Cheng, Qing, Perer, Holstein, Wu and Zhu}]{cheng2022child}
Cheng HF, Stapleton L, Kawakami A, Sivaraman V, Cheng Y, Qing D, Perer A, Holstein K, Wu ZS and Zhu H (2022) How child welfare workers reduce racial disparities in algorithmic decisions.
\newblock In: \emph{Proceedings of the 2022 CHI Conference on Human Factors in Computing Systems}. pp. 1--22.

\bibitem[{Cho et~al.(2023)Cho, Toffey, Silva, Shalev, Safford, Phillips, Lee, Wiggins, Kozlov, Tsui et~al.}]{cho2023care}
Cho J, Toffey B, Silva AF, Shalev A, Safford MM, Phillips E, Lee A, Wiggins F, Kozlov E, Tsui EK et~al. (2023) To care for them, we need to take care of ourselves: {A} qualitative study on the health of home health aides.
\newblock \emph{Health services research} 58(3): 697--704.

\bibitem[{Connacher(2023)}]{aidelow}
Connacher C (2023) Direct care worker pay and benefits are low despite high demand for services.
\newblock \urlprefix\url{https://www.americanprogress.org/article/direct-care-worker-pay-and-benefits-are-low-despite-high-demand-for-services/}.

\bibitem[{Coston et~al.(2023)Coston, Kawakami, Zhu, Holstein and Heidari}]{coston2022validity}
Coston A, Kawakami A, Zhu H, Holstein K and Heidari H (2023) A validity perspective on evaluating the justified use of data-driven decision-making algorithms.
\newblock In: \emph{2023 IEEE conference on secure and trustworthy machine learning (SaTML)}. IEEE, pp. 690--704.

\bibitem[{Cotter et~al.(2001)Cotter, Hermsen, Ovadia and Vanneman}]{cotter2001glass}
Cotter DA, Hermsen JM, Ovadia S and Vanneman R (2001) The glass ceiling effect.
\newblock \emph{Social forces} 80(2): 655--681.

\bibitem[{{Council on Social Work Education}(2023)}]{socialworknums}
{Council on Social Work Education} (2023) 2022 - 2023 {Statistics} on social work education in the {United} {States}.

\bibitem[{Delgado et~al.(2023)Delgado, Yang, Madaio and Yang}]{delgado2023participatory}
Delgado F, Yang S, Madaio M and Yang Q (2023) The participatory turn in {AI} design: Theoretical foundations and the current state of practice.
\newblock In: \emph{Proceedings of the 3rd ACM Conference on Equity and Access in Algorithms, Mechanisms, and Optimization}. pp. 1--23.

\bibitem[{Dressel and Farid(2018)}]{dressel2018accuracy}
Dressel J and Farid H (2018) The accuracy, fairness, and limits of predicting recidivism.
\newblock \emph{Science Advances} .

\bibitem[{Elwyn et~al.(2013)Elwyn, Scholl, Tietbohl, Mann, Edwards, Clay, L{\'e}gar{\'e}, Weijden, Lewis, Wexler et~al.}]{elwyn2013many}
Elwyn G, Scholl I, Tietbohl C, Mann M, Edwards AG, Clay C, L{\'e}gar{\'e} F, Weijden Tvd, Lewis CL, Wexler RM et~al. (2013) “{Many} miles to go…”: {A} systematic review of the implementation of patient decision support interventions into routine clinical practice.
\newblock \emph{BMC medical informatics and decision making} 13(Suppl 2): S14.

\bibitem[{England(2005)}]{england2005emerging}
England P (2005) Emerging theories of care work.
\newblock \emph{Annual Review of Sociology} 31(1): 381--399.

\bibitem[{Enyedy(2014)}]{enyedy2014personalized}
Enyedy N (2014) Personalized instruction: {New} interest, old rhetoric, limited results, and the need for a new direction for computer-mediated learning .

\bibitem[{Esposito et~al.(2016)Esposito, Fiorini, Limosani, Bonaccorsi, Manzi, Cavallo and Dario}]{esposito2016supporting}
Esposito R, Fiorini L, Limosani R, Bonaccorsi M, Manzi A, Cavallo F and Dario P (2016) Supporting active and healthy aging with advanced robotics integrated in smart environment.
\newblock In: \emph{Optimizing Assistive Technologies for Aging Populations}.

\bibitem[{Filippi et~al.(2023)Filippi, Bann{\`o} and Trento}]{filippi2023automation}
Filippi E, Bann{\`o} M and Trento S (2023) Automation technologies and the risk of substitution of women: Can gender equality in the institutional context reduce the risk?
\newblock \emph{Technological Forecasting and Social Change} 191: 122528.

\bibitem[{Flores-Robles and Gantman(2024)}]{flores2023notions}
Flores-Robles G and Gantman AP (2024) Notions of care labor are antithetical to profitable labor.
\newblock \emph{Psychology of Women Quarterly} 48(4): 475--490.

\bibitem[{Forsythe(1993)}]{forsythe1993construction}
Forsythe DE (1993) The construction of work in artificial intelligence.
\newblock \emph{Science, technology, \& human values} 18(4): 460--479.

\bibitem[{Foucault(2023)}]{foucault2023discipline}
Foucault M (2023) Discipline and punish.
\newblock In: \emph{Social Theory Re-Wired}.

\bibitem[{Fox et~al.(2023)Fox, Shorey, Kang, Montiel~Valle and Rodriguez}]{fox2023patchwork}
Fox SE, Shorey S, Kang EY, Montiel~Valle D and Rodriguez E (2023) Patchwork: the hidden, human labor of {AI} integration within essential work.
\newblock \emph{Proceedings of the ACM on Human-Computer Interaction} 7(CSCW1): 1--20.

\bibitem[{Fox et~al.(2019)Fox, Sobel and Rosner}]{fox2019managerial}
Fox SE, Sobel K and Rosner DK (2019) Managerial visions: {Stories} of upgrading and maintaining the public restroom with {IoT}.
\newblock In: \emph{Proceedings of the 2019 CHI Conference on Human Factors in Computing Systems}. pp. 1--15.

\bibitem[{Gerchick et~al.(2023)Gerchick, Jegede, Shah, Gutierrez, Beiers, Shemtov, Xu, Samant and Horowitz}]{gerchick2023devil}
Gerchick M, Jegede T, Shah T, Gutierrez A, Beiers S, Shemtov N, Xu K, Samant A and Horowitz A (2023) The devil is in the details: {Interrogating} values embedded in the allegheny family screening tool.
\newblock In: \emph{Proceedings of the 2023 ACM Conference on Fairness, Accountability, and Transparency}. pp. 1292--1310.

\bibitem[{Gilman(2018)}]{automatedimm}
Gilman J (2018) Automated immunity.
\newblock \urlprefix\url{https://brownpoliticalreview.org/2018/02/automated-immunity/}.

\bibitem[{Haraway(2013)}]{haraway2013situated}
Haraway D (2013) Situated knowledges: The science question in feminism and the privilege of partial perspective 1.
\newblock In: \emph{Women, science, and technology}. Routledge, pp. 455--472.

\bibitem[{Hendry and Wu(2004)}]{hendry2004care}
Hendry J and Wu Y (2004) \emph{The care of the elderly in {Japan}}.
\newblock Routledge.

\bibitem[{Hittle et~al.(2016)Hittle, Agbonifo, Suarez, Davis and Ballard}]{hittle2016complexity}
Hittle B, Agbonifo N, Suarez R, Davis KG and Ballard T (2016) Complexity of occupational exposures for home health-care workers: {Nurses} vs. home health aides.
\newblock \emph{Journal of nursing management} 24(8): 1071--1079.

\bibitem[{Hoffmann(2021)}]{hoffmann2021terms}
Hoffmann AL (2021) Terms of inclusion: Data, discourse, violence.
\newblock \emph{New Media \& Society} 23(12): 3539--3556.

\bibitem[{Holmes(2023)}]{holmes2023unintended}
Holmes W (2023) The unintended consequences of artificial intelligence and education .

\bibitem[{Holstein et~al.(2017)Holstein, McLaren and Aleven}]{holstein2017intelligent}
Holstein K, McLaren BM and Aleven V (2017) Intelligent tutors as teachers' aides: {Exploring} teacher needs for real-time analytics in blended classrooms.
\newblock In: \emph{Proceedings of the seventh international learning analytics \& knowledge conference}. pp. 257--266.

\bibitem[{Holstein et~al.(2018)Holstein, McLaren and Aleven}]{Holstein2018}
Holstein K, McLaren BM and Aleven V (2018) {Student learning benefits of a mixed-reality teacher awareness tool in AI-enhanced classrooms}.
\newblock In: \emph{International Conference on Artificial Intelligence in Education}.
\newblock ISBN 9783319938431, pp. 154--168.

\bibitem[{Holstein et~al.(2019)Holstein, Mclaren and Aleven}]{holstein2019}
Holstein K, Mclaren BM and Aleven V (2019) {Co-designing a real-time classroom orchestration tool to support teacher–{AI} complementarity}.
\newblock \emph{Journal of Learning Analytics} .

\bibitem[{Holten~M{\o}ller et~al.(2020)Holten~M{\o}ller, Shklovski and Hildebrandt}]{holten2020}
Holten~M{\o}ller N, Shklovski I and Hildebrandt TT (2020) Shifting concepts of value: Designing algorithmic decision-support systems for public services.
\newblock In: \emph{Proceedings of the 11th Nordic Conference on Human-Computer Interaction: Shaping Experiences, Shaping Society}. pp. 1--12.

\bibitem[{Huising et~al.(2025)Huising, Elmholdt and M{\"a}kinen}]{huising2025introduction}
Huising R, Elmholdt KT and M{\"a}kinen E (2025) Introduction: The changing constitution and ecology of expertise.
\newblock \emph{Research in the Sociology of Organizations} .

\bibitem[{Irani and Silberman(2013)}]{irani2013turkopticon}
Irani LC and Silberman MS (2013) Turkopticon: Interrupting worker invisibility in amazon mechanical turk.
\newblock In: \emph{Proceedings of the SIGCHI conference on human factors in computing systems}. pp. 611--620.

\bibitem[{J{\o}rgensen and Nissen(2022)}]{jorgensen2022making}
J{\o}rgensen AM and Nissen MA (2022) Making sense of decision support systems: {Rationales}, translations and potentials for critical reflections on the reality of child protection.
\newblock \emph{Big Data \& Society} 9(2): 20539517221125163.

\bibitem[{June(2022)}]{june2022oregon}
June N (2022) Oregon is dropping an artificial intelligence tool used in child welfare system.
\newblock \emph{Link: https://www. npr. org/2022/06/02/1102661376/oregon-drops-artificial-intelligence-child-abuse-cases} .

\bibitem[{Kawakami et~al.(2024{\natexlab{a}})Kawakami, Coston, Heidari, Holstein and Zhu}]{kawakamistudying2024}
Kawakami A, Coston A, Heidari H, Holstein K and Zhu H (2024{\natexlab{a}}) Studying up public sector {AI}: How networks of power relations shape agency decisions around {AI} design and use.
\newblock \emph{Proceedings of the ACM on Human-Computer Interaction} 8(CSCW2): 1--24.

\bibitem[{Kawakami et~al.(2024{\natexlab{b}})Kawakami, Coston, Zhu, Heidari and Holstein}]{kawakami2024situate}
Kawakami A, Coston A, Zhu H, Heidari H and Holstein K (2024{\natexlab{b}}) The {Situate} {AI} {Guidebook}: Co-designing a toolkit to support multi-stakeholder early-stage deliberations around public sector {AI} proposals : 1--22.

\bibitem[{Kawakami et~al.(2022{\natexlab{a}})Kawakami, Sivaraman, Cheng, Stapleton, Cheng, Qing, Perer, Wu, Zhu and Holstein}]{kawakami2022improving}
Kawakami A, Sivaraman V, Cheng HF, Stapleton L, Cheng Y, Qing D, Perer A, Wu ZS, Zhu H and Holstein K (2022{\natexlab{a}}) Improving human-{AI} partnerships in child welfare: Understanding worker practices, challenges, and desires for algorithmic decision support.
\newblock In: \emph{Proceedings of the 2022 CHI Conference on Human Factors in Computing Systems}. pp. 1--18.

\bibitem[{Kawakami et~al.(2022{\natexlab{b}})Kawakami, Sivaraman, Stapleton, Cheng, Perer, Wu, Zhu and Holstein}]{kawakami2022care}
Kawakami A, Sivaraman V, Stapleton L, Cheng HF, Perer A, Wu ZS, Zhu H and Holstein K (2022{\natexlab{b}}) “{Why} do i care what’s similar?” {Probing} challenges in {AI}-assisted child welfare decision-making through worker-{AI} interface design concepts.
\newblock In: \emph{Proceedings of the 2022 ACM Designing Interactive Systems Conference}. pp. 454--470.

\bibitem[{Kocon(2017)}]{teachingFem}
Kocon A (2017) Is teaching undervalued because it’s “women’s work”?
\newblock \urlprefix\url{https://tntp.org/blog/is-teaching-undervalued-because-its-womens-work/}.

\bibitem[{Kuo et~al.(2022)Kuo, Cho, Olaye, Delgado, Dell and Sterling}]{kuo2022understanding}
Kuo EFC, Cho J, Olaye I, Delgado D, Dell N and Sterling MR (2022) Understanding the technological landscape of home health aides: {Scoping} literature review and a landscape analysis of existing mhealth apps.
\newblock \emph{Journal of Medical Internet Research} 24(11): e39997.

\bibitem[{Lebovitz et~al.(2021)Lebovitz, Levina and Lifshitz-Assaf}]{lebovitz2021ai}
Lebovitz S, Levina N and Lifshitz-Assaf H (2021) Is {AI} ground truth really true? {The} dangers of training and evaluating ai tools based on experts’ know-what.
\newblock \emph{MIS quarterly} 45(3): 1501--1526.

\bibitem[{Lebovitz et~al.(2022)Lebovitz, Lifshitz-Assaf and Levina}]{lebovitz2022engage}
Lebovitz S, Lifshitz-Assaf H and Levina N (2022) To engage or not to engage with ai for critical judgments: How professionals deal with opacity when using ai for medical diagnosis.
\newblock \emph{Organization science} 33(1): 126--148.

\bibitem[{Mateescu and Elish(2019)}]{mateescu2019ai}
Mateescu A and Elish M (2019) {AI} in context: {The} labor of integrating new technologies .

\bibitem[{Migda(2015)}]{newdeal}
Migda A (2015) The legacies of slavery and jim crow live on with exclusion of home health care workers from fair labor laws.

\bibitem[{Ming et~al.(2023)Ming, Kuo, Go, Tseng, Kallas, Vashistha, Sterling and Dell}]{ming2023go}
Ming J, Kuo E, Go K, Tseng E, Kallas J, Vashistha A, Sterling M and Dell N (2023) ``{I} go beyond and beyond": {E}xamining the invisible work of home health aides.
\newblock \emph{Proceedings of the ACM on Human-Computer Interaction} 7(CSCW1): 1--21.

\bibitem[{Mod{\'e}n et~al.(2021)Mod{\'e}n, Tallvid, Lundin and Lindstr{\"o}m}]{utterberg2021intelligent}
Mod{\'e}n MU, Tallvid M, Lundin J and Lindstr{\"o}m B (2021) Intelligent tutoring systems: Why teachers abandoned a technology aimed at automating teaching processes .

\bibitem[{Muller and Kuhn(1993)}]{muller1993participatory}
Muller MJ and Kuhn S (1993) Participatory design.
\newblock \emph{CACM} .

\bibitem[{Orlikowski(1992)}]{orlikowski1992learning}
Orlikowski WJ (1992) Learning from notes: Organizational issues in groupware implementation.
\newblock In: \emph{Proceedings of the 1992 ACM conference on Computer-supported cooperative work}. pp. 362--369.

\bibitem[{Osterman(2018)}]{osterman2018improving}
Osterman P (2018) Improving long-term care by finally respecting home-care aides.
\newblock \emph{Hastings Center Report} 48: S67--S70.

\bibitem[{Osterman(2019)}]{osterman2019improving}
Osterman P (2019) Improving job quality for direct care workers.
\newblock \emph{Economic Development Quarterly} 33(2): 151--156.

\bibitem[{Pruss(2023)}]{pruss2023ghosting}
Pruss D (2023) Ghosting the machine: Judicial resistance to a recidivism risk assessment instrument.
\newblock In: \emph{Proceedings of the 2023 ACM conference on fairness, accountability, and transparency}. pp. 312--323.

\bibitem[{Ratner and Elmholdt(2023)}]{ratner2023algorithmic}
Ratner HF and Elmholdt K (2023) Algorithmic constructions of risk: Anticipating uncertain futures in child protection services.
\newblock \emph{Big Data \& Society} 10(2): 20539517231186120.

\bibitem[{Roll and Wylie(2016)}]{roll2016evolution}
Roll I and Wylie R (2016) Evolution and revolution in artificial intelligence in education.
\newblock \emph{International journal of artificial intelligence in education} 26(2): 582--599.

\bibitem[{Rury(1991)}]{rury1991education}
Rury JL (1991) \emph{Education and Women's Work: {Female} Schooling and the Division of Labor in Urban America, 1870-1930. SUNY Series on Women and Work.}
\newblock ERIC.

\bibitem[{Samant et~al.(2021)Samant, Horowitz, Xu and Beiers}]{samant2021family}
Samant A, Horowitz A, Xu K and Beiers S (2021) Family surveillance by algorithm: The rapidly spreading tools few have heard of.

\bibitem[{Saxena et~al.(2021)Saxena, Badillo-Urquiola, Wisniewski and Guha}]{saxena2021framework}
Saxena D, Badillo-Urquiola K, Wisniewski PJ and Guha S (2021) A framework of high-stakes algorithmic decision-making for the public sector developed through a case study of child-welfare.
\newblock \emph{Proceedings of the ACM on Human-Computer Interaction} 5(CSCW2): 1--41.

\bibitem[{Schwandt and Gates(2018)}]{schwandt2018case}
Schwandt TA and Gates EF (2018) Case study methodology.
\newblock \emph{The SAGE handbook of qualitative research} 5: 600--630.

\bibitem[{Selwyn(2016)}]{edtechkey}
Selwyn N (2016) \emph{Education and Technology: Key Issues and Debates}.
\newblock Bloomsbury Publishing.

\bibitem[{Sendak et~al.(2020)Sendak, Ratliff, Sarro, Alderton, Futoma, Gao, Nichols, Revoir, Yashar, Miller et~al.}]{sendak2020real}
Sendak MP, Ratliff W, Sarro D, Alderton E, Futoma J, Gao M, Nichols M, Revoir M, Yashar F, Miller C et~al. (2020) Real-world integration of a sepsis deep learning technology into routine clinical care: {Implementation} study.
\newblock \emph{JMIR medical informatics} 8(7): e15182.

\bibitem[{Sharkey and Wood(2014)}]{sharkey2014paro}
Sharkey A and Wood N (2014) The {Paro} seal robot: {Demeaning} or enabling.
\newblock In: \emph{Proceedings of AISB}, volume~36. p. 2014.

\bibitem[{Sloane et~al.(2022)Sloane, Moss, Awomolo and Forlano}]{sloane2022participation}
Sloane M, Moss E, Awomolo O and Forlano L (2022) Participation is not a design fix for machine learning.
\newblock In: \emph{Proceedings of the 2nd ACM Conference on Equity and Access in Algorithms, Mechanisms, and Optimization}. pp. 1--6.

\bibitem[{Smith(2022)}]{teachingWomen}
Smith K (2022) Teaching in the light of women's history.
\newblock \urlprefix\url{https://www.facinghistory.org/ideas-week/teaching-light-womens-history}.

\bibitem[{Solano-Kamaiko et~al.(2025)Solano-Kamaiko, Tan, Ming, Avgar, Vashistha, Sterling and Dell}]{solano2025running}
Solano-Kamaiko IR, Tan M, Ming J, Avgar AC, Vashistha A, Sterling M and Dell N (2025) ``{W}ho is running it?'' {T}owards equitable {AI} deployment in home care work.
\newblock In: \emph{Proceedings of the 2025 CHI Conference on Human Factors in Computing Systems}. pp. 1--19.

\bibitem[{Spektor et~al.(2023)Spektor, Fox, Awumey, Begleiter, Kulkarni, Stringam, Riordan, Rho, Akridge and Forlizzi}]{spektor2023charting}
Spektor F, Fox SE, Awumey E, Begleiter B, Kulkarni C, Stringam B, Riordan CA, Rho HJ, Akridge H and Forlizzi J (2023) Charting the automation of hospitality: {An} interdisciplinary literature review examining the evolution of frontline service work in the face of algorithmic management.
\newblock \emph{Proceedings of the ACM on Human-Computer Interaction} 7(CSCW1): 1--20.

\bibitem[{Spiel et~al.(2020)Spiel, Brul{\'e}, Frauenberger, Bailley and Fitzpatrick}]{spiel2020details}
Spiel K, Brul{\'e} E, Frauenberger C, Bailley G and Fitzpatrick G (2020) In the details: the micro-ethics of negotiations and in-situ judgements in participatory design with marginalised children.
\newblock \emph{CoDesign} 16(1): 45--65.

\bibitem[{Spinuzzi(2002)}]{spinuzzi2002scandinavian}
Spinuzzi C (2002) A scandinavian challenge, a {US} response: {Methodological} assumptions in scandinavian and {US} prototyping approaches.
\newblock In: \emph{Proceedings of the 20th annual international conference on Computer documentation}. pp. 208--215.

\bibitem[{Star(1999)}]{star1999ethnography}
Star SL (1999) The ethnography of infrastructure.
\newblock \emph{American behavioral scientist} 43(3): 377--391.

\bibitem[{Star and Strauss(1999)}]{star1999layers}
Star SL and Strauss A (1999) Layers of silence, arenas of voice: The ecology of visible and invisible work.
\newblock \emph{Computer supported cooperative work (CSCW)} 8(1): 9--30.

\bibitem[{Steiner and Woo(2021)}]{steiner2021job}
Steiner ED and Woo A (2021) Job-related stress threatens the teacher supply.
\newblock \emph{Rand Corporation} 25: 08--1.

\bibitem[{Suchman(1993)}]{suchman1993categories}
Suchman L (1993) Do categories have politics? {The} language/action perspective reconsidered.
\newblock \emph{Computer supported cooperative work (CSCW)} 2(3): 177--190.

\bibitem[{Suchman(1995)}]{suchman1995making}
Suchman L (1995) Making work visible.
\newblock \emph{Communications of the ACM} 38(9): 56--64.

\bibitem[{Suchman(2011)}]{suchman2011anthropological}
Suchman L (2011) Anthropological relocations and the limits of design.
\newblock \emph{Annual review of anthropology} 40(1): 1--18.

\bibitem[{Suchman(1987)}]{suchman1987plans}
Suchman LA (1987) \emph{Plans and situated actions: The problem of human-machine communication}.
\newblock Cambridge.

\bibitem[{Taie and Lewis(2022)}]{taie2022characteristics}
Taie S and Lewis L (2022) Characteristics of 2020-21 public and private k-12 school teachers in the {United} {States}: Results from the national teacher and principal survey.
\newblock \emph{National Center for Education Statistics} .

\bibitem[{Tseng et~al.(2020)Tseng, Okeke, Sterling and Dell}]{tseng2020we}
Tseng E, Okeke F, Sterling M and Dell N (2020) "{We} can learn. why not?" {Designing} technologies to engender equity for home health aides.
\newblock In: \emph{Proceedings of the 2020 CHI conference on human factors in computing systems}. pp. 1--14.

\bibitem[{Vaithianathan et~al.(2017)Vaithianathan, Putnam-Hornstein, Jiang, Nand and Maloney}]{AFSTdocumentation}
Vaithianathan R, Putnam-Hornstein E, Jiang N, Nand P and Maloney T (2017) Developing predictive models to support child maltreatment hotline screening decisions: Allegheny county methodology and implementation.
\newblock \emph{Center for Social data Analytics} .

\bibitem[{Wang et~al.(2024)Wang, Kapoor, Barocas and Narayanan}]{wang2022against}
Wang A, Kapoor S, Barocas S and Narayanan A (2024) Against predictive optimization: {On} the legitimacy of decision-making algorithms that optimize predictive accuracy.
\newblock \emph{ACM Journal on Responsible Computing} 1(1): 1--45.

\bibitem[{Watters(2015)}]{edhistfuture}
Watters A (2015) The history of the future of education.

\bibitem[{Wright(2018)}]{wright2018tactile}
Wright J (2018) Tactile care, mechanical hugs: Japanese caregivers and robotic lifting devices.
\newblock \emph{Asian Anthropology} 17(1): 24--39.

\bibitem[{Wright(2019)}]{wright2019robots}
Wright J (2019) Robots vs migrants? {Reconfiguring} the future of {Japanese} institutional eldercare.
\newblock \emph{Critical Asian Studies} 51(3): 331--354.

\bibitem[{Wyatt and Altman(1995)}]{wyatt1995commentary}
Wyatt JC and Altman DG (1995) Commentary: Prognostic models: clinically useful or quickly forgotten?
\newblock \emph{Bmj} 311(7019): 1539--1541.

\bibitem[{Yang et~al.(2019)Yang, Steinfeld and Zimmerman}]{yang2019unremarkable}
Yang Q, Steinfeld A and Zimmerman J (2019) Unremarkable {AI}: Fitting intelligent decision support into critical, clinical decision-making processes.
\newblock In: \emph{Proceedings of the 2019 CHI conference on human factors in computing systems}. pp. 1--11.

\bibitem[{Yang et~al.(2016)Yang, Zimmerman, Steinfeld, Carey and Antaki}]{yang2016investigating}
Yang Q, Zimmerman J, Steinfeld A, Carey L and Antaki JF (2016) Investigating the heart pump implant decision process: opportunities for decision support tools to help.
\newblock In: \emph{Proceedings of the 2016 CHI Conference on Human Factors in Computing Systems}. pp. 4477--4488.

\bibitem[{Yildirim et~al.(2023)Yildirim, Oh, Sayar, Brand, Challa, Turri, Crosby~Walton, Wong, Forlizzi, McCann et~al.}]{yildirim2023creating}
Yildirim N, Oh C, Sayar D, Brand K, Challa S, Turri V, Crosby~Walton N, Wong AE, Forlizzi J, McCann J et~al. (2023) Creating design resources to scaffold the ideation of ai concepts.
\newblock In: \emph{Proceedings of the 2023 ACM Designing Interactive Systems Conference}. pp. 2326--2346.

\bibitem[{York(2022)}]{socialwork}
York CHN (2022) There is no such thing as women’s work: The gender of labor and the devaluation of social workers.
\newblock \urlprefix\url{https://ny.covenanthouse.org/there-is-no-such-thing-as-womens-work-the-gender-of-labor-and-the-devaluation-of-social-workers/}.

\bibitem[{Zhang et~al.(2024)Zhang, Rana, Boltz, Dubal and Lee}]{zhang2024data}
Zhang A, Rana R, Boltz A, Dubal V and Lee MK (2024) Data probes as boundary objects for technology policy design: {Demystifying} technology for policymakers and aligning stakeholder objectives in rideshare gig work.
\newblock In: \emph{Proceedings of the 2024 CHI Conference on Human Factors in Computing Systems}. pp. 1--21.

\end{thebibliography}

\end{document}